\documentclass[11pt,a4paper]{article}
\pdfoutput=1
\usepackage{epsfig}
\usepackage{jheppub}

\newcommand{\be}{\begin{equation}}
\newcommand{\ee}{\end{equation}}
\newcommand{\bear}{\begin{eqnarray}}
\newcommand{\eear}{\end{eqnarray}}
\newcommand{\ba}{\begin{array}}
\newcommand{\ea}{\end{array}}
\newcommand{\lae}{\begin{array}{c}\,\sim\vspace{-21pt}\\<
\end{array}}

\title{Full-hierarchy Quiver Theories of  Electroweak Symmetry 
Breaking   and Fermion Masses}

\author[a]{Gustavo Burdman}
\author[a]{Nayara Fonseca}
\author[a]{Leonardo de Lima}
\affiliation[a]{
Instituto de F\'{i}sica, Universidade de S\~{a}o Paulo,
\\ R. do Mat\~{a}o 187, S\~{a}o Paulo, SP 05508-900, Brazil }
\emailAdd{burdman@if.usp.br, nayara.focs@gmail.com, leonardodelima@gmail.com}

\abstract{
We consider quiver theories in four dimensions with a large ultra-violet
cutoff. These theories require that an ordered set of vacuum expectation
values for the link fields develops dynamically and can be obtained
from the coarse deconstruction of extra-dimensional theories in an AdS
background.  These full-hierarchy quiver theories form a large class
which include AdS$_5$ models as a limit, but which have a distinctive
phenomenology.  
As an example, in this paper we show that 
fermions can be introduced in a way that can at the
same time generate the fermion mass hierarchy and have flavor
violation consistent with experimental bounds, when the mass scale of
the color-octect gauge excitation is above $3~{\rm TeV}$.  
We also show that electroweak precision constraints are 
satisfied by this mass scale, without the need to extend the gauge
sector to protect against custodial violation.
}


\begin{document}
 
\maketitle


\section{Introduction} \setcounter{equation}{0}
\label{intro}
Despite the continued experimental success~\cite{pdg} of the standard model (SM),
several fundamental questions remain unanswered in particle physics.
Among these are the origin of electroweak symmetry breaking and
fermions masses. In the SM, the Higgs sector is composed of a scalar
doublet, which results in gauge bosons and fermion masses as it
acquires a vacuum expectation value (VEV). 
This symmetry breaking sector of the SM leaves a remnant state, the
Higgs boson, which may have been recently observed at the LHC~\cite{higgs1}. More
importantly, even if the Higgs boson is observed it remains unclear
what keeps the weak scale --in the SM determined by the scale at which
the Higgs VEV develops-- from running away to the cutoff of the
theory, presumably a much higher ultra-violet  (UV) scale such as the
Planck mass, $M_{P}$.  The problem of keeping the weak scale separated
from the UV cutoff in a theory with an elementary Higgs boson is what we call the hierarchy problem. 

It is possible to protect the Higgs mass from the quadratic
divergences that would drive it and the weak scale to the UV cutoff by
making use of symmetries. Such is the case with weak-scale
supersymmetric models~\cite{susy}. However, the origin of the supersymmetry
breaking scale leading to the now stable weak scale, becomes the new
unknown. In addition, the generation of fermion masses posses a
problem typically leading to large flavor violation and CP
asymmetries. 
Alternatively, it is possible to naturally generate a large
weak-scale/UV-scale hierarchy by assuming a new assimptotically-free
interaction getting strong at the TeV scale, such as in Technicolor
theories~\cite{tc}.  These scenarios also have a problem with large
flavor violation since in order to accommodate fermion masses the
interaction has to be embedded in extended gauge sectors. 
Furthermore, the need to add new chiral fermions has an unavoidable
impact in electroweak precision constraints, particularly the S
parameter. 

More recently, a way of solving the hierarchy problem   by using
a curved extra dimension in an anti--de~Sitter (AdS) background has
been formulated~\cite{rs1,rsbulk,gp00}. The underlying assumption in
AdS$_5$ theories is that
they are in some way dual to 4D strongly coupled sectors near
conformality,  such as walking technicolor (WTC)~\cite{tc}. Much like
in 4D strongly coupled theories, these models tend to have a large S
parameter, which can however be controlled by moderately increasing
the Kaluza-Klein (KK) mass scale. Initially, it appeared that the fermion
mass hierarchy could be naturally explained by the localization of the fermion
zero-modes in the extra dimension, controlled by parameters of order
one~\cite{rsbulk,gp00}, all  while not incurring in too large tree-level
flavor violation.  However, the flavor-violating contributions of KK
gluons to kaon observables pushes  (in the most generic scenario)
the KK mass scale up to a few tens of TeV, resulting in
considerable fine-tuning~\cite{csabafv}. Thus, these 5D constructions
expected to be dual to 4D strongly coupled sectors,  present 
problems similar to  the ones plaguing the original TC-ETC models. 

In this paper we present a set of theories that are designed to
generate a large hierarchy between the weak scale and a large UV
scale. These purely four-dimensional models consist of gauge groups
joined by sigma models, the so-called quiver theories, and they are
cousins of AdS$_5$ theories in that they are related to them by the 
procedure called deconstruction~\cite{decon1}.  They 
are characterized by highly ordered vacuum expectation values (VEVs)  of the sigma
models, starting just below the desired UV scale and going all the way
to the weak scale.
The limit of  large number of gauge groups and large value of the UV
VEV, corresponds to recovering AdS$_5$ theories and we will call it
the continuum limit. However, the four-dimensional quiver theories can
be quite
different from AdS$_5$, particularly when we consider the opposite
limit, which we can call coarse discretization, and that corresponds
to using only a few gauge groups. 
 Thus, we are interested in working  far from the continuum
limit to build our models, so that these quiver theories cannot have
an interpretation as a low-energy AdS$_5$ model that can be obtained
by truncating the Kaluza-Klein towers~\cite{kkcut}. As we will show
below, the result of a coarse discretization is very different from
this procedure since, for instance, it gives very different couplings of the excited
states to zero-mode fermions, resulting in a distinct phenomenology.
We will show that quiver theories with few gauge groups or ``sites''
can generate a large hierarchy of scales, as well as the fermion mass
hierarchy just as AdS$_5$ models, while not generating too large
flavor violation. This particular advantage, together with the fact
that these theories are fundamentally different from the warped extra
dimension models, makes a full exploration of full-hierarchy quiver theories an
interesting proposition. 

At a more fundamental level, we see that AdS$_5$ theories are only a
small fraction of a much larger set of theories capable of generating
large hierarchies. Formulating these vast set of theories, what we
call Full-hierarchy Quiver Theories (FHQT), with AdS$_5$ as  a
limiting case for large number of groups, allows us to generalize the
good features of AdS$_5$ theories to their coarse FHQT cousins. Since
these are quantitatively
different from  AdS$_5$ we expect important phenomenological
distinctions between them and FHQT models with a sufficiently small number of sites. 
In order to start studying these differences, we will build coarse
quiver theories very similar to bulk AdS$_5$ models 
and explicitly show that
they can have potentially much less flavor violation than them. 
Here we will consider models that in the continuum limit would result
in an infrared (IR) localized Higgs. However, it is possible to obtain
a dynamical Higgs in FHQT very much in the same way that composite
Higgs models emerge in AdS$_5$. We leave this to a separate companion 
publication~\cite{higgs}. In general,  it is possible to reproduce any
model in the AdS$_5$ limit of FHQT, so we could study it in the coarse
limit.  Thus, besides the issue of flavor violation with a localized
Higgs, or reproducing Gauge-Higgs unification in a coarse theory, we
can consider various other issues in this general framework. 
For instance, issues from how  the conformal behavior of the AdS$_5$ theories
manifests itself in the quiver theories, to the existence of a
strongly coupled ``dual''  which  the four-dimensional quiver theories
correspond to, should be addressed in the future. For now, we will
concentrate on the phenomenological aspects related to model building
the mechanism for electroweak symmetry breaking (EWSB) and 
fermion masses. But more theoretical ramifications may and should be
addressed concerning the use of FHQT and their behavior as quantum
field theories. 

The models will have a stabilized
gauge hierarchy and a natural origin for the fermion hierarchy. 
Just as in warped extra dimensions, the origin of the fermion mass
hierarchy is {\em a priori}  independent of the stabilization of the
Higgs sector. The gauge hierarchy problem is solved as long as the
Higgs is IR-localized. IR localization of the Higgs can be achieved
through specific mechanisms. For the purpose of this paper we assume
it will be fully localized in the so called IR site, which in the
continuum limit would correspond to localizing the Higgs in the IR
brane. Dynamical Higgs localization is left for a separate
work~\cite{higgs}, and can be achieved in a way very similar to
composite models~\cite{chm} in the continuum limit.
In this setup, we will consider flavor by introducing fermion
localization in theory space, and will study the resulting flavor
violation phenomenology in coarse FHQT. We will show that in these
cases it is
possible to have very little flavor violation, unlike in the continuum
limit. We will also consider the electroweak bounds on coarse
FHQT. In the AdS$_5$ case, these impose an extension in the
electroweak gauge sector due to the presence of large isospin
violation. We will show that in general FHQT do not need 
custodial protection to have sufficiently small contributions to the
$T$ parameter.  

There are several papers in the literature making use of deconstruction techniques 
to obtain models of electroweak symmetry breaking and/or flavor. 
For instance, in Ref.~\cite{threesite} a 3-site Higgsless model is
proposed, and its flavor structure studied in ~\cite{threesiteflav}. 
In Ref.~\cite{rediflavsym} the flavor physics of composite Higgs
models like those of ~\cite{chm} is explored using a 2-site setup with
flavor symmetries. In Ref.~\cite{redi4dcomp} the Higgs sector of this
setup is considered, whereas 2 and 3-site models with custodial
protection are considered in Ref.~\cite{dischm}. A more general
approach to a 4D pNGB Higgs is presented in Ref.~\cite{chmgen},
although it is still tied to the coset $SO(5)/SO(4)$ presented in \cite{chm}.
 In our work we considered a larger number of sites
in order to explore the feasibility of the non-hierarchical primordial
Yukawa couplings. In doing so, we will see that our results, if viewed
from a 2-site model perspective, encode the resulting effective flavor
symmetries of the fully-deconstructed AdS$_5$ theory. In addition, we
will see that it is not necessary to consider custodial  protection in
the gauge sector of the model in order to avoid too large a tree-level
contribution to the $T$ parameter. As a result, the minimal composite
Higgs model will have a smaller symmetry that in the papers above.
Then  the FHQT constructions bring
a new perspective to both the flavor and electroweak sectors.

The rest of the paper is organized as follows: in Section~\ref{sec1}
we present the FHQT and their relation to
a coarse deconstruction  of AdS$_5$ models. In Section~\ref{sec2} we
show how fermions are included and how the fermion mass hierarchy can
be obtained naturally. We study flavor violation in
Section~\ref{sec3}, where we show that it is possible to obtain small
enough violation to accommodate all experimental bounds, while keeping
the new physics mass scale close to the TeV. The electroweak precision
bounds are considered in~\ref{ewpc}. 
We finally conclude in Section~\ref{conc}.

\section{Full-hierarchy Quiver Theories}
\label{sec1}
\begin{figure}
\begin{center}
\includegraphics[scale=0.8]{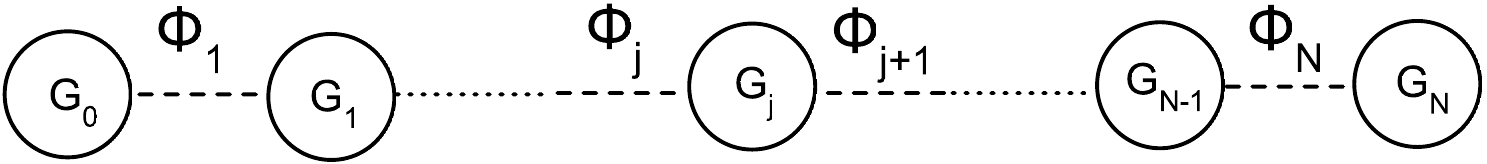}
\caption{Quiver diagram for the theory described by (\ref{s1})..} 
\label{f:1}
\end{center}
\end{figure}

In this section we describe the basics of FHQT. As an example let us consider the product 
gauge group $G_0\times G_1\times\cdots G_j\times G_{j+1}\cdots G_N$.
In addition, we have a set of scalar link fields $\Phi_j$, with
$j=1~{\rm to~} N$, transforming as bi-fundamentals under $G_{j-1}\times G_{j}$. 
The action for the theory is 
\be
S = \int d^4x \left\{ -\,\sum_{j=0}^N\,\frac{1}{2g_j^2} \,Tr\left[ F_{\mu\nu}^{(j)}
    F^{\mu\nu (j)}\right]  + \sum_{j=1}^N \,Tr\left[
    (D_\mu\Phi_j)^\dagger D^{\mu} \Phi_j\right] -V(\Phi_j)
+\dots
\right\}
\label{s1}
\ee
where the traces are over the groups'  generators, and the dots at the
end correspond to terms involving fermions and  will be discussed
in the next section. We assume that the potentials
for the link fields give them a vacuum expectation value (VEV) which breaks $G_{j-1} \times G_j$ down to the
diagonal group, and result in non-linear sigma models for the
$\Phi$'s
\be
\Phi_j = \frac{v_j}{\sqrt{2}}\, e^{i\sqrt{2} \pi_j^a \hat t^a/v_j}~,
\label{phis}
\ee  
 where the $\hat t^a$'s are the broken generators, the $\pi_j^a$ the
 Nambu-Goldstone Bosons (NGB); and $v_j$ are the VEVs of the link
 fields. We will consider here the situation where the VEVs are
 ordered in such a way that $v_1\dots  > v_j\dots > v_N$. 
We parametrize the ordering by defining the VEVs as 
\be
v_j \equiv v q^j~,
\label{vjdef}
\ee
where $0<q<1$ is a dimensionless constant, and $v$ is a UV mass scale
that can be regarded as the UV cutoff.  We will also assume that the
all the gauge groups are identical and that their 
gauge couplings satisfy
\be
g_0(v)=g_1(v_1)=\dots=g_j(v_j) = g_{j+1}(v_{j+1})=\dots \equiv g~.
\label{gaugeequal}
\ee
The model can be illustrated
by the quiver diagram of Figure~\ref{f:1}. 

This purely 4D theory can be  obtained from deconstructing an
extra-dimensional theory in an AdS$_5$ background~\cite{starfleet,decads5,tools}.
Discretizing a 5D gauge theory in an AdS$_5$ background by a discrete
interval $1/gv$ in $N$ intervals results in the action (\ref{s1}),
with the appropriate identification of the 5D gauge coupling, plus the
matching 
\be
q \leftrightarrow  e^{-k/gv}~.
\label{qmatching}
\ee
However, in order for the 4D theory defined by (\ref{s1}) to remain a good
description of the continuum 5D theory, the  AdS$_5$ curvature
should satisfy  $k<v$, or $q$ close to 1. When this is satisfied, getting closer to the
continuum limit by increasing the number of sites $N$ guarantees an
increasing similarity with the 5D theory~\cite{tools}. 
For instance, generating the hierarchy between the Planck
and the weak scales while satisfying $k<v$ requires typically that 
$N>35$, which results in a low energy theory very close to the
continuum one.  Under these conditions, 4D theories with $k<v$
are just discrete descriptions of the AdS$_5$ theory. 

On the other hand,  if we consider (\ref{s1}) as just a 4D theory, we
are free to make use of values of $q$ far from what would constitute
the continuum 5D limit, i.e. $q\ll 1$. In these theories it will be
possible to obtain a large hierarchy of scales with smaller values of
$N$, as low as just a few.  For instance, if $v\lae M_{P}$ and
$v_N\simeq O(1)~$TeV, then   we can write
\be
q  = 10^{-16/N} ~.
\label{pl2v}
\ee
For instance, for $N=4$ we have $q=10^{-4}$, very far from the
continuum limit. The theories resulting in these region of the
parameters of the action in (\ref{s1}) will have a   very different behavior than a
mere discretization of  AdS$_5$. Their spectrum and its properties,
such as couplings to SM matter, differ significantly and therefore
they merit a detailed study.  

There are several aspects of FHQT worth exploring. Regarding their use
to build models of EWSB, the
most urgent appears to be the modeling of the Higgs sector leading to
EWSB, and the generation of fermion masses. We consider the dynamical
origin of the Higgs sector in a separate publication~\cite{higgs},
where the Higgs is a remnant pseudo-NGB (pNGB). For this paper, we
concentrate on the issue of fermion masses and assume a very simple
Higgs sector, one that  captures the essential features of the Yukawa
interactions in these models. 
Our aim is to explore the consequences of naturally generating the
fermion mass hierarchy in FHQT. Specifically, we want to know if is
possible to build 
models with acceptable levels of flavor violation. This is important
in light of the great difficulties encountered in AdS$_5$ models
regarding this issue~\cite{adsfv}. It is also generally of great
interest to explore new models of the fermion hierarchy  and their
flavor-violating effects. 

\section{Fermion Localization in Quiver Space} 
\label{sec2}
In this section we incorporate fermions to the FHQT. The main goal is
to model fermion masses in the context of Higgs sector models that
solve the hierarchy problem within the framework of FHQT. 

\begin{figure}
\begin{center}
\includegraphics[scale=0.8]{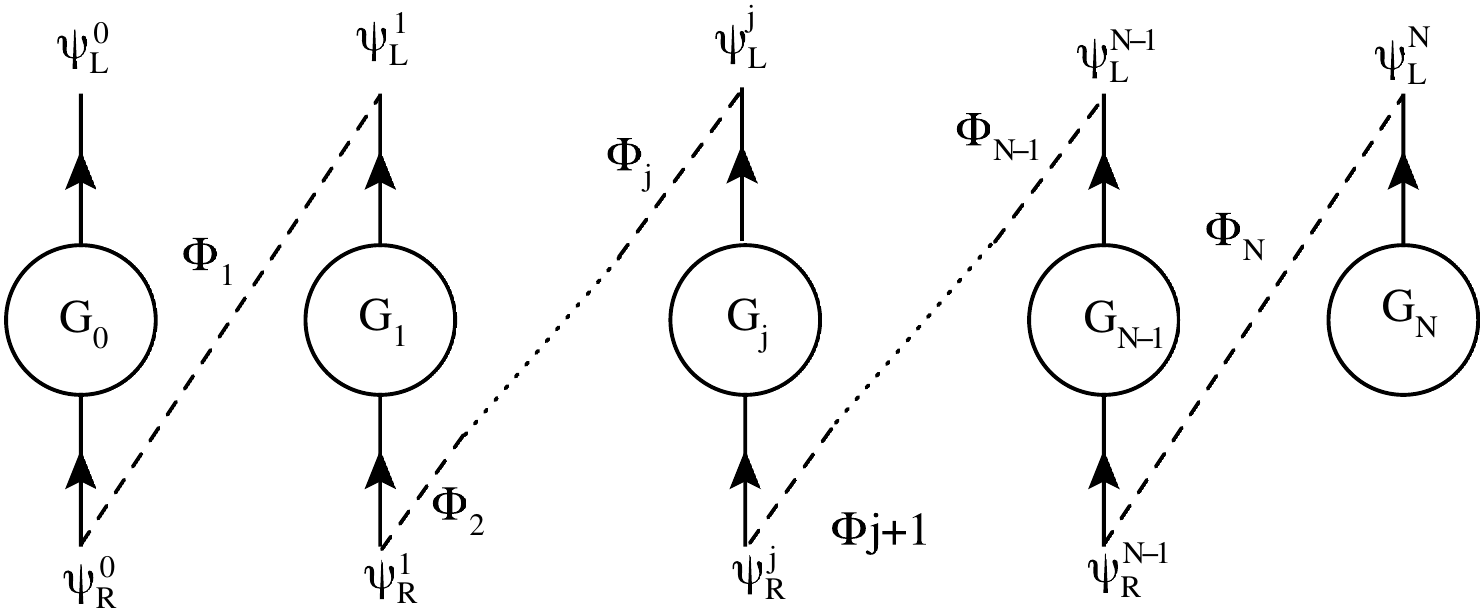}
\caption{Quiver diagram for the theory described by (\ref{sf}), for a
  spectrum with a left-handed zero mode.} 
\label{quiver2}
\end{center}
\end{figure}

We consider vector-like fermions $\psi^{j}$ transforming in the
fundamental representation of the groups $G_j$.  
The action of (\ref{s1}) is then enlarged by the fermion action given by 
\be
S_f = \int d^4x \sum_{j=0}^N \left\{ \bar\psi_L^{j} i\hspace*{-0.1cm}\not\hspace*{-0.1cm}D_j \psi_L^{j}
+ \bar\psi_R^{j} i\hspace*{-0.1cm}\not\hspace*{-0.1cm}D_j \psi_R^{j} -
(\mu_j \bar\psi_L^{j}\psi_R^{j} +
\lambda_j\bar\psi_R^{j-1}\Phi_j\psi_L^{j} + {\rm h.c.} )~,
\right\}
\label{sf}
\ee 
which is represented by the quiver diagram of Figure~\ref{quiver2}.
The vector-like masses $\mu_j$ preserve the gauge symmetries. 
The Yukawa term is invariant since the links transform as $\Phi_j\to
g_{j-1}\Phi_j g^\dagger_j$. The Yukawa couplings are allowed to be site-dependent, which is the 
most general situation in the 4D theory. If one wanted to match to the 
continuum limit of the AdS$_5$ theory we should take them to be
universal, as shown in Ref~\cite{bbh}. 
In the unitary gauge we make the replacement $\Phi_j \to
v_j/\sqrt{2}$, which leads to a non-diagonal mass matrix for the
fermions.  We diagonalize to the mass eigenstate basis through the 
unitary transformations 
\be
\psi_{L,R}^{j} = \sum_{n=0}^N h_{L,R}^{j,n} \,\chi_{L,R}^{(n)} ~,
\label{rotation}   
\ee 
where the $\chi_{L,R}^{(n)}$ are the mass eigenstates. Imposing the
equations of motion, results in the elements of
the rotation matrices satisfying the equations~\cite{bbh}
\bear
(\mu_j^2+\frac{\lambda_j^2v_j^2}{2}-m_n^2)\,h_L^{j,n} -
\frac{\lambda_jv_j}{\sqrt{2}}\mu_{j-1} \,h_L^{j-1,n} -
\frac{\lambda_{j+1}v_{j+1}}{\sqrt{2}}\mu_jh_L^{j+1,n} &=& 0\\
(\mu_j^2+\frac{\lambda_{j+1}^2v_{j+1}^2}{2}-m_n^2)\,h_R^{j,n} -
\frac{\lambda_jv_j}{\sqrt{2}}\mu_{j}\, h_R^{j-1,n} -
\frac{\lambda_{j+1}v_{j+1}}{\sqrt{2}}\mu_{j+1}\,h_R^{j+1,n} &=& 0
\eear
where $m_n$ is the mass of the mass eigenstate $\chi_{L,R}^{(n)}$. The
solutions of these equations can be obtained~\cite{tools} and in the
continuum limit would match to the solutions for the wave-functions of
the Kaluza-Klein fermions in the AdS$_5$~\cite{bbh}. But here we stay
far from the continuum. 

We are interested in studying the  fermion zero-modes.  These satisfy the
simple equations of motion  
\be
\mu_jh_L^{j,0}  + \frac{\lambda_{j+1}}{\sqrt{2}} v_{j+1} h_L^{j+1,0} = 0~,
\label{lhzm}
\ee
for the left-handed zero mode, and 
\be
\mu_j\,h_R^{j,0}  + \frac{\lambda_{j}}{\sqrt{2}} v_{j} h_R^{j-1,0}= 0~,
\label{rhzm}
\ee
for the right-handed zero mode. We can define the localization parameter
$\nu_L$ for the left-handed zero mode  by 
\be
\sqrt{2}\,\frac{\mu_j}{v\,\lambda_{j+1}} \equiv - q^{j+1+\nu_L}~,
\label{nuldef}
\ee
and then consistently identify the localization parameter $\nu_R$  for
a right-handed zero mode by
\be
\sqrt{2}\,\frac{\mu_j}{v\,\lambda_j} =- q^{j+1+\nu_R} ~,
\ee
Then, we can see that 
\be
\frac{h_L^{j+1,0}}{h_L^{j,0}} = q^{\nu_L}~,\qquad 
\frac{h_R^{j,0}}{h_R^{j-1,0}} = q^{-(1+\nu_R)} ~. 
\ee
Thus, we have traded the ratio of vector masses to Yukawa couplings
for a parameter ($\nu_L$ or $\nu_R$) that will determine how much of
each fermion in the quiver diagram the zero-mode fermion contains. 
We can then write 
\be
h_{L,R}^{j,0}  = z^j_{L,R}\,h_{L,R}^{0,0} ~,
\label{lrjmode}
\ee
where we have defined 
\be
z_L\equiv q^{\nu_L}~,\qquad  z_R \equiv q^{-(1+\nu_R)} ~.
\label{zlrdef}
\ee
On the other hand, the normalization conditions require that
\be
\sum_{j=0}^N |h_{L,R}^{j,0}|^2 = 1~,
\ee
which we use to obtain
\be
h_{L,R}^{0,0} = \sqrt{\frac{1-z_{L,R}^2}{1-z_{L,R}^{2(N+1)}}}~, 
\ee

We can use these results to compute the couplings of the fermion zero
mode to various states. Here we will consider first the Yukawa
couplings to the Higgs. As it is shown in Ref.~\cite{higgs}, in order
to solve the hierarchy problem, the Higgs must be coupled mostly to
the gauge groups that survive at lower energies, particularly
$G_N$. This is the analog of having a Higgs localized close to the IR
brane in AdS$_5$ models in 5D. For simplicity we will assume that the
Higgs only transforms under $G_N$ (``localized'' at the end of the
quiver).  In Ref.~\cite{higgs} we show dynamical mechanisms to do just
that. Here we are interested in the Yukawa couplings and their general
behavior. With such a ``N-Localized'' Higgs,  the Yukawa coupling of a
given zero-mode fermion is defined by
\be
{\cal L} = - {\cal Y} \bar\psi_L^N H \psi_R^N + h.c.~ = -{\cal Y} h_L^{*N,0}
h_R^{N,0} \bar\chi_L^{(0)} H \chi_R^{(0)} + h.c.~, 
\ee 
where the primordial Yukawa ${\cal Y}$ is assumed to be an $O(1)$ number. 
The resulting effective Yukawa coupling of the fermion zero-mode is 
\be
Y = {\cal Y} \, z_L^N\, z_R^N\, h_L^{0,0}\,h_R^{0,0}
~.\label{yukawa4d}
\ee
Thus, we see that zero-mode fermions with large components of their
eigenstate wave-functions coming from $\psi_{L,R}^{i}$ with $i$
close or equal to $N$, will have unsuppressed Yukawa couplings with
the ``N-localized'' Higgs. On the other hand, zero-mode fermions with
wave-functions built mostly of $\psi_{L,R}^{i}$ with $i$ closer to
$0$ will have largely suppressed Yukawa couplings in
(\ref{yukawa4d}). We can then build theories of flavor by choosing the 
quiver-diagram  localization of the zero-mode fermion, much in the same
way it is done in AdS$_5$ theories.
Fermion-mass hierarchies can be
built by appropriately choosing the localization parameters for the
fermions in the quiver theory, $\nu_L$ and $\nu_R$ for each
generation.  This can be achieved with quiver theories with a small
number of sites, such as 4 or 5, very far from the continuum limit. 

One potential worry is the presence of flavor changing neutral
currents (FCNCs) at tree level, induced by the non-universal couplings
of massive gauge bosons to the zero-mode fermions that necessarily
appear as a consequence of the different localization in the quiver
diagram. In fact, these FCNCs are a very important problem in AdS$_5$
theories and result in the most stringent of bounds~\cite{csabafv}.
In the next section, we address this issue in FHQT and show that in
them it is possible to build the SM fermion mass hierarchy without large
FCNCs at tree level. In particular, we show that is possible to find
solutions where FCNCs are nearly absent in the down-type quark sector,
which is typically the most binding. 

\section{Mass Hierarchies and Flavor Violation}
\label{sec3} 
In order to study flavor violation in FHQT, we first need to compute
the couplings of massive gauge bosons to zero-mode fermions. We first
notice that the wave-function of a zero-mode fermion can be written as
\be
\chi^{(0)}_{L,R} = \sum_{j=0}^N\,h_{L,R}^{*j,0}\,\psi_{L,R}^j~,
\label{zmfermion}
\ee
in terms of the quiver fermions.  Likewise, and assuming a generic
gauge group in the sites of the quiver diagram, the mass-eigenstates
of the gauge bosons can be written in terms of the quiver gauge bosons
as
\be
A_\mu^{(n)} = \sum_{j=0}^N\, f^*_{j,n}\,A^j_\mu~,
\label{nmode}
\ee
with $f_{j,n}$ the coefficient linking the gauge boson in site $j$
with the mass-eigenstate $n$ in the rotation to mass eigenstates.
We are interested in  obtaining the coupling of the  $n=1$ to the
zero-mode fermions since this state gives the largest FCNCs
effect. We can compute this coupling
\be
g^{01}_{L,R}\, \bar\chi_{L,R}^{(0)}\gamma^\mu A_\mu^{(1)} \chi_{L,R}^{(0)}
~,
\label{g01def}
\ee
where we assumed that group generators are absorbed in the definition
of the gauge fields, and we obtain
\be
g^{01}_{L,R} = \sum_{j=0}^N\,
g_j\,\left|h_{L,R}^{j,0}\right|^2\,f_{j,1}~,
\label{g01}
\ee
where $g_j$ are the gauge couplings associated to the group $G_j$  in
the quiver. As mentioned above, we will assume $g_j=g$ for all
$j$. To be more precise, we actually mean to say that 
$g_j=g(v_j)$, with $g(\mu)$ the same {\em running} coupling for all
gauge groups in the quiver. 

 For fixed values of $N$ the coefficients $f_{j,1}$ can
be obtained by diagonalizing the gauge boson mass matrix~\cite{tools,bbh}. Then, we can
obtain the coupling of zero-mode fermions to the first excited state
of the gauge bosons, normalized by the gauge coupling $g$. 
These couplings are of great interest for various reasons. For
instance, the  s-channel production of the first-excited states of the
gauge bosons is determined by them, so they would play a central role
in the collider searches for these theories.
But they could also lead to tree-level flavor violation since their
couplings to the SM fermions are generally not universal. 
The couplings in (\ref{g01}) depend on the localization parameters
$\nu_L$ and $\nu_R$, which are to be chosen appropriately to get the
correct zero-mode masses as well as mixings, as mentioned in the
previous section.

\subsection{Localization and Flavor Violation}
\label{locvsfv}

\begin{figure} 
\begin{center}
\includegraphics[scale=0.9]{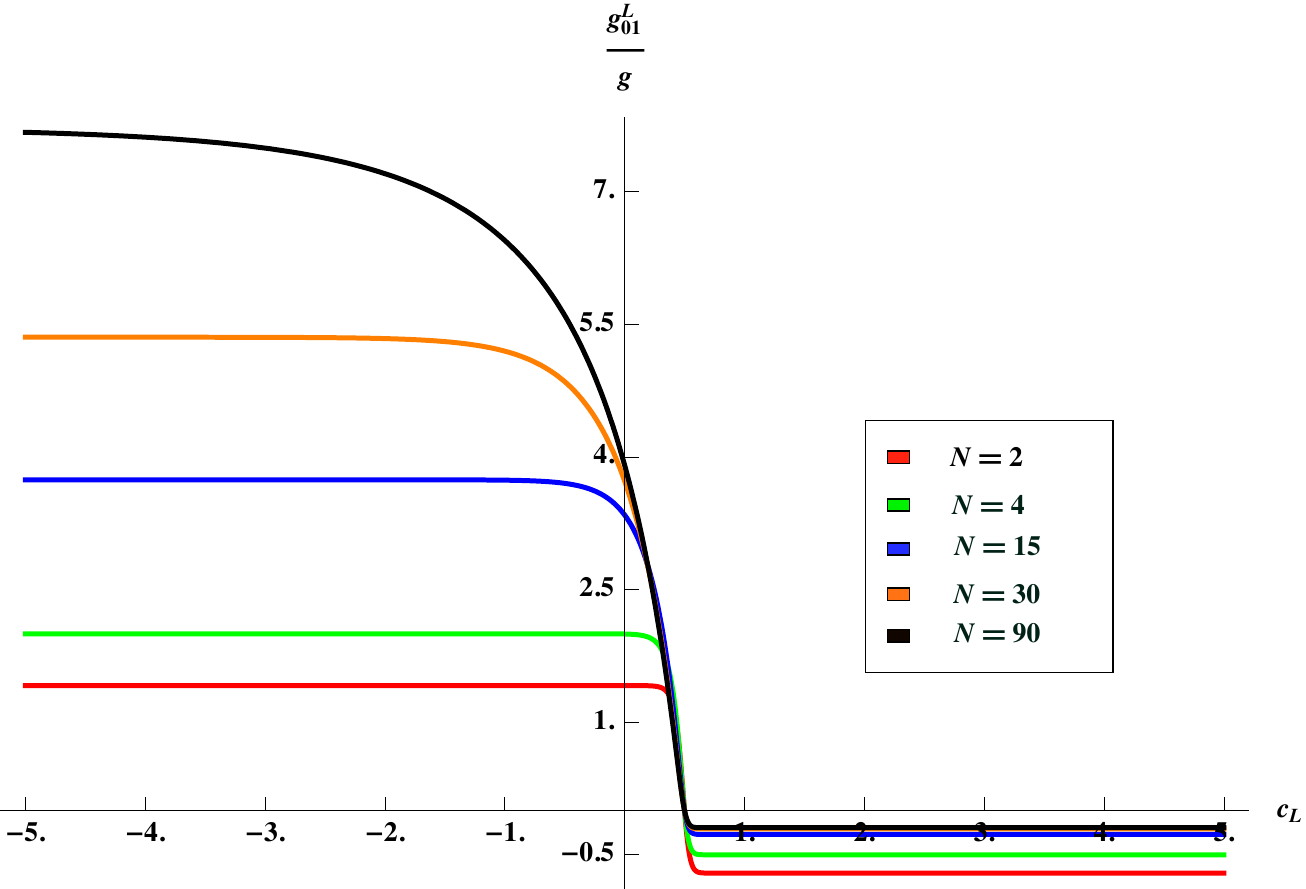}
\caption{
Couplings of a left-handed zero-mode fermion to the first excited
state of a gauge boson, normalized to the zero-mode gauge boson
coupling, as a function of the localization parameter $c_L$.
For the left side of the plot and starting from the bottom: $N=2$,
$N=4$, $N=15$, $N=30$ and $N=90$. }
\label{g1vsn}
\end{center}
\end{figure}

Here we consider the coupling of the zero-mode fermions to the first
excited state of a gluon. In Figure~\ref{g1vsn} we show the coupling
of a left-handed zero-mode fermion as a function of the localization
parameter redefined as  $c_L\equiv \nu_L+1/2$, and for various values
of the number of sites in the quiver, N. We can see that as N
increases  from small values, corresponding to coarse discretization,
to large values nearing the continuum limit, the coupling goes to its
continuum limit, as it can be verified by comparing the $N=90$ case
with the results in Ref.~\cite{gp00}. 
We can see then that these couplings of great phenomenological
importance are quite different far from the continuum. The same can be
done for the right-handed couplings.

\begin{figure}
\begin{center}
\includegraphics[scale=0.9]{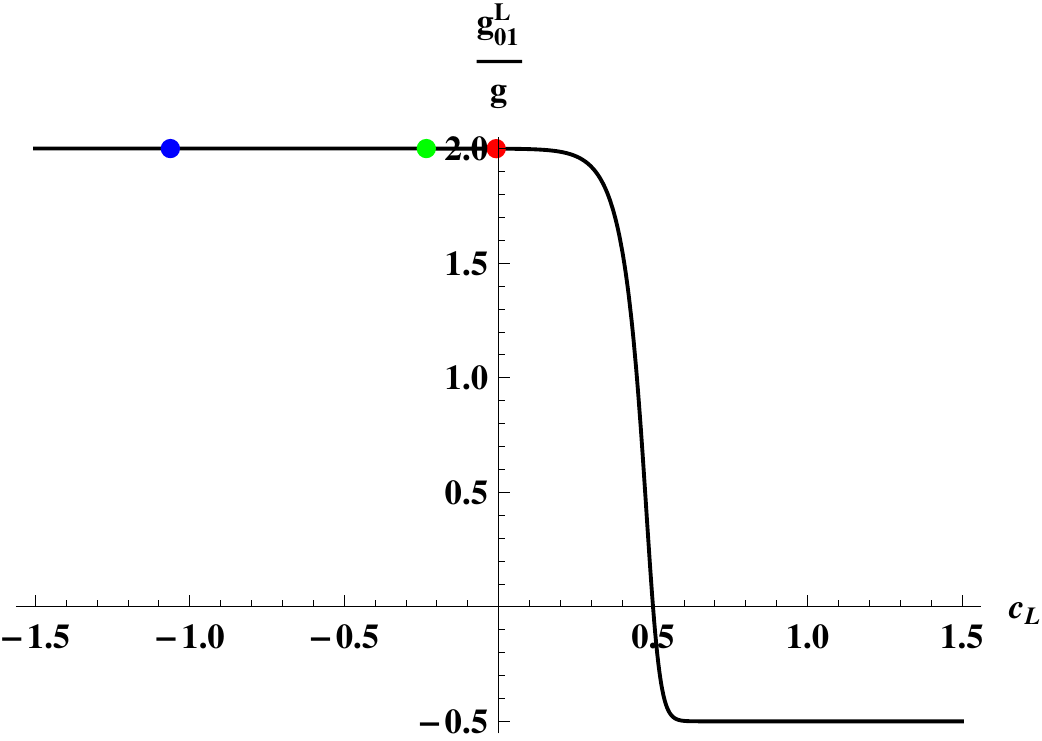}
\caption{
Couplings of the left-handed zero-mode quarks to the
first excited state of the gauge boson, in units of the zero-mode
gauge coupling, vs. the localization parameter $c_L$ defined in the
text (solid line). The dots correspond to the localization of the
solution called {\bf case A}. }
\label{fig:2}
\end{center}
\end{figure}

\begin{figure}
\begin{center}
\includegraphics[scale=0.9]{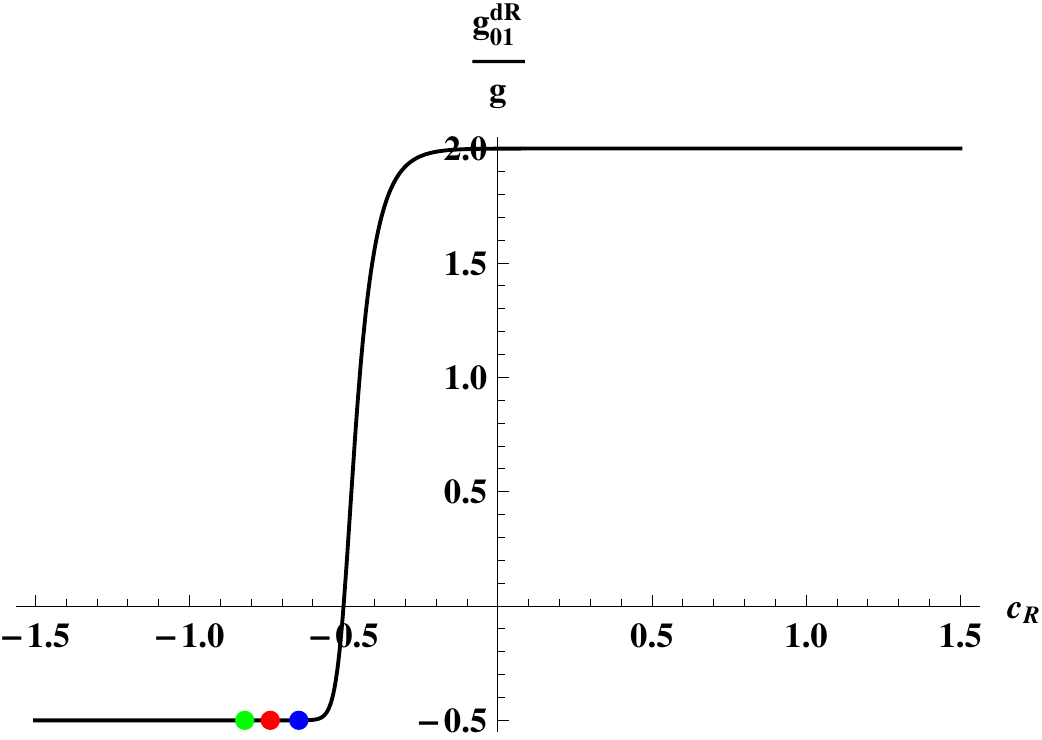}
\caption{
Couplings of the right-handed zero-mode down quarks to the
first excited state of the gauge boson, in units of the zero-mode
gauge coupling, vs. the localization parameter $c_R$ defined in the
text (solid line). The dots correspond to the localization of the
solution called {\bf case A}. }
\label{fig:3}
\end{center}
\end{figure}

\begin{figure} 
\begin{center}
\includegraphics[scale=0.9]{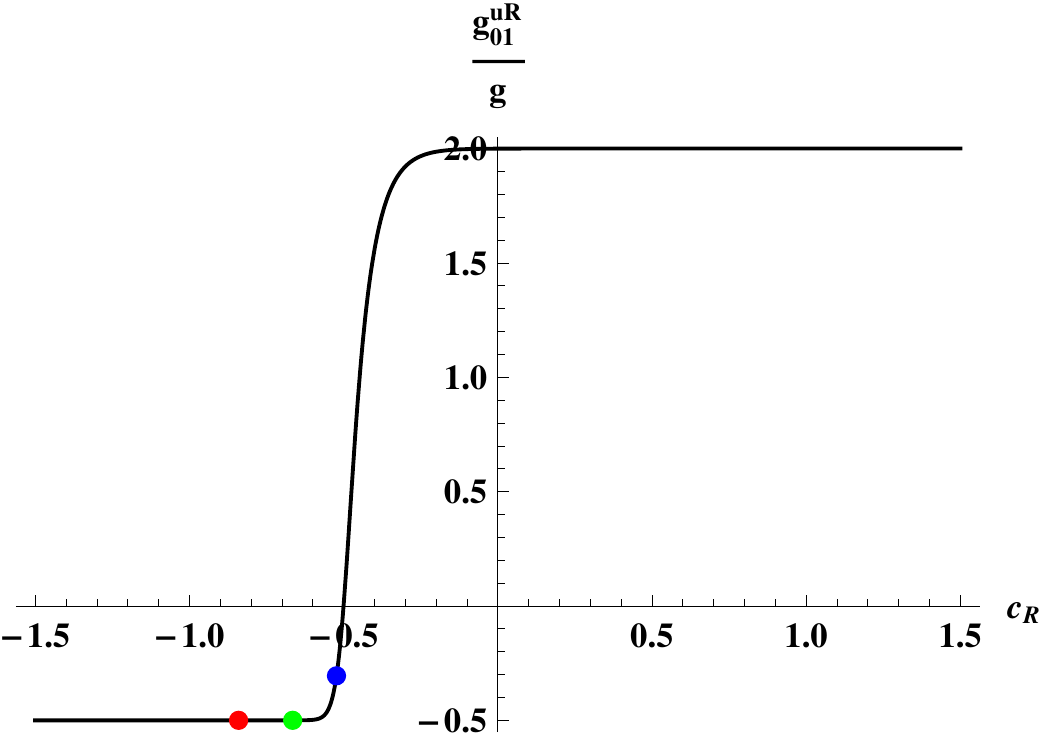}
\caption{
Couplings of the right-handed zero mode up quarks to the
first excited state of the gauge boson, in units of the zero-mode
gauge coupling, vs. the localization parameter $c_R$ defined in the
text (solid line). The dots correspond to the localization of the
solution called {\bf case A}. }
\label{fig:4}
\end{center}
\end{figure}

As an example, we study these couplings for the  $N=4$ case, i.e five sites. We
compute the coupling of zero-mode fermions to the first gauge
excitation as a function of the localization parameter for the
zero-mode fermion, $\nu_{L,R}$. The case of a left-handed zero mode
fermion is shown in the solid
line in Figure ~\ref{fig:2}.
We observe that there are two plateaus: one above $c_L>1/2$, the other
for $c_L<0.25$.  The transition region is rather small. Thus, a given
solution for the $c_L$'s (i.e. for the zero-mode fermion masses and
mixings) such that they are on either plateau, will have effectively
very small or negligible tree-level flavor violation. 
We can do the same for the right-handed down and up zero-mode
fermions. The corresponding couplings to the first-excited gauge boson
state are plotted in the solid lines of Figures~\ref{fig:3} and \ref{fig:4}. 

Thus,  we want to find solutions for the localization parameters
$c_{L,R}^i$'s, where the $i=1,2,3$ denotes generation, which lay
mostly on the plateaus in order to minimize flavor violation.
We will show two such examples. In the first case, {\bf case A}, we minimize the
amount of flavor violation necessary to obtain the correct masses and
mixings in the quark sector. This is achieved by localizing the
right-handed sector as close to the  UV plateau as possible, while the
left-handed  quark sector is localized towards the IR plateau.  
In the second case, {\bf case B}, the right-handed down sector remains in
the UV plateau, whereas the left-handed sector and most of the
right-handed up sector is also there, but with small amounts of flavor
violation. 

\paragraph{Case A:}
In Figure~\ref{fig:2}, we plot the coupling of the first excited state
of a gauge boson (normalized to its zero-mode coupling)  as a function
of the localization parameters in the quiver diagram.   
The dots show a solution for the quark sector, that results in the correct
masses and mixings. In this case, the left-handed quarks are localized
in the IR plateau. Although
their localizations in the quiver imply different Yukawa couplings,
the will have nearly universal couplings to the first excited state 
of the gauge bosons, resulting in very little or no tree-level flavor violation. 
The solution for the localization coefficients
$c^i_{L,R}$ are color coded. We see in Figure~\ref{fig:2}  that the couplings of the
left-handed doublet zero mode quark to the first gauge excitation are
universal, so they will not  result in tree-level flavor violation. 
The same can be concluded by observing the couplings of the down-type
right handed zero-mode quarks, Figure~\ref{fig:3}. Thus, we see that
this solution of the $N=4$ example{ \em does not result in tree-level flavor violation
in the down-quark sector}. This should be compared with the situation
in RS models in an AdS$_5$ background, where flavor violation at tree
level is unavoidable in the down sector, leading to very stringent
bounds on the scale of the gauge excitations~\cite{adsfv}. 
Finally, the right-handed up-sector couplings to the first gauge boson
excitation are shown in Figure~\ref{fig:4}. We see that it is not
possible to have universal couplings as in the right-handed down
sector. This is due to the need for the right-handed top to be 
closer to the IR in order to obtain the correct top quark mass. This
will lead to the leading source of tree-level flavor violation in
this case, which will be in the up sector in observables such as
$D^0\bar D^0 $ mixing. In the next subsection we show that this effect
is very small and quite compatible with experiment. 

As
we increase $N$,  the curves in Figures~\ref{fig:2}, \ref{fig:3} and
\ref{fig:4}  will tend to their analogous in the continuum AdS$_5$
limit~\cite{gp00}. Thus, we see that, although  FHQT share some of the
features of AdS$_5$ models, they behave in a different way when it comes to
the amount of flavor violation induced. This difference is one of the key
points that make FHQT more viable phenomenologically than models based
on  warped extra dimensions. 

\paragraph{Case B}  

\begin{figure} 
\begin{center}
\includegraphics[scale=0.9]{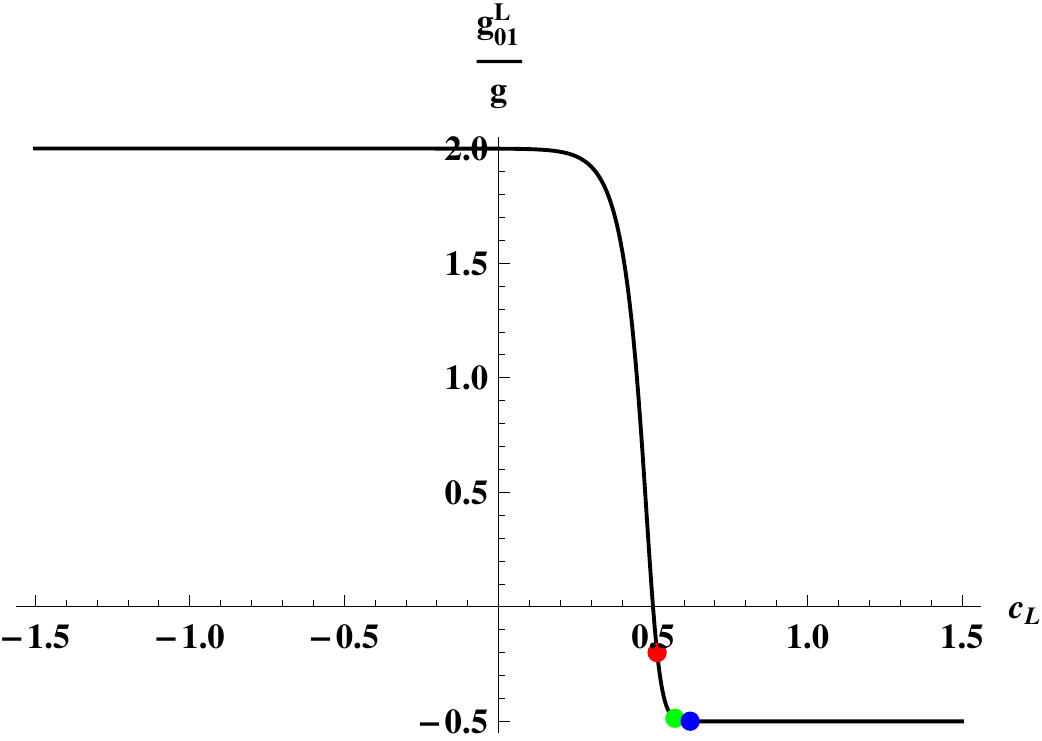}
\caption{
Couplings of the left-handed zero-mode quarks to the
first excited state of the gauge boson, in units of the zero-mode
gauge coupling, vs. the localization parameter $c_L$ defined in the
text (solid line). The dots correspond to the localization of the
solution called {\bf case B}. }
\label{fig:5}
\end{center}
\end{figure}

\begin{figure} 
\begin{center}
\includegraphics[scale=0.9]{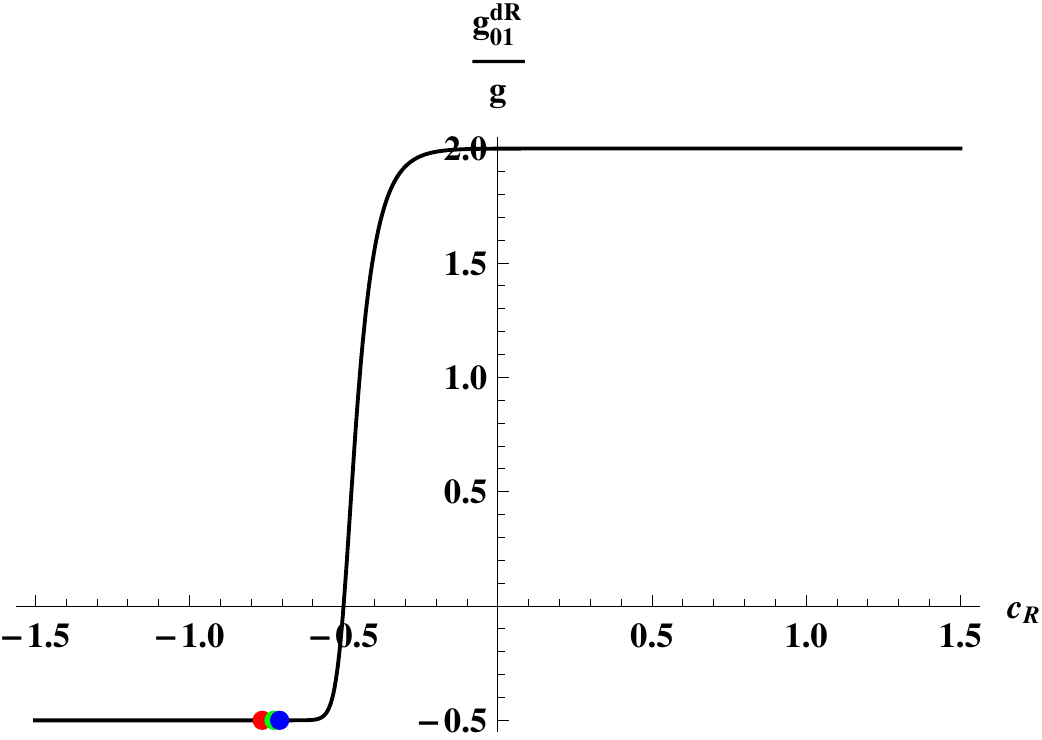}
\caption{
Couplings of the right-handed zero-mode down quarks to the
first excited state of the gauge boson, in units of the zero-mode
gauge coupling, vs. the localization parameter $c_R$ defined in the
text (solid line). The dots correspond to the localization of the
solution called {\bf case B}. }
\label{fig:6}
\end{center}
\end{figure}

\begin{figure} 
\begin{center}
\includegraphics[scale=0.9]{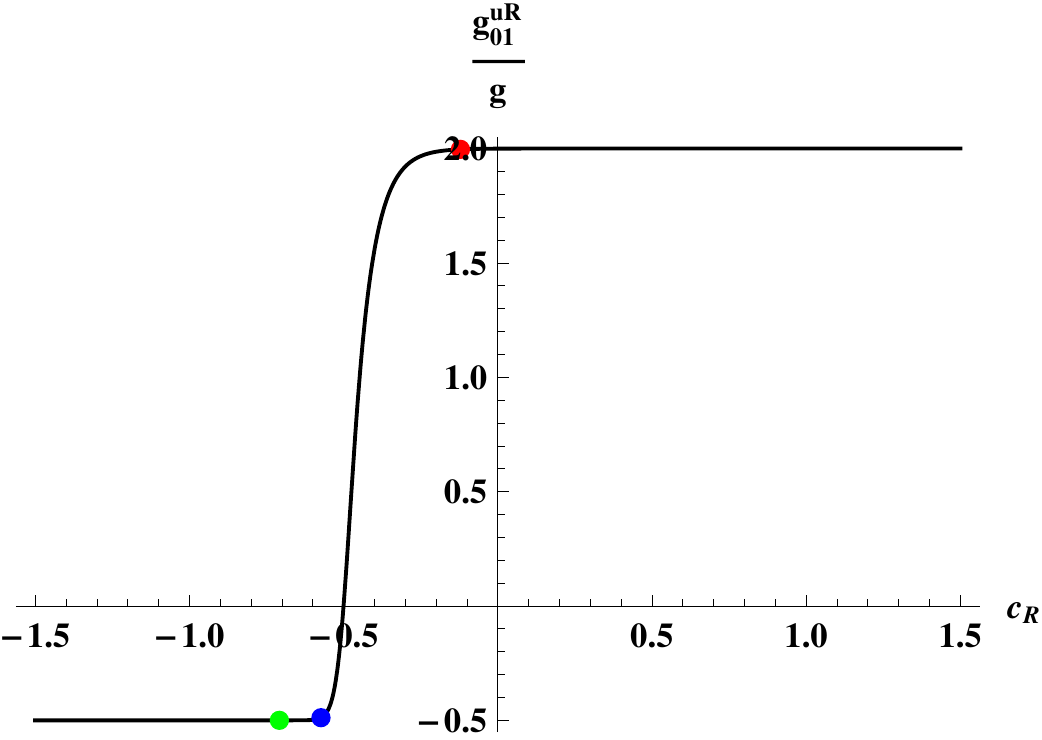}
\caption{
Couplings of the right-handed zero-mode up quarks to the
first excited state of the gauge boson, in units of the zero-mode
gauge coupling, vs. the localization parameter $c_R$ defined in the
text (solid line). The dots correspond to the localization of the
solution called {\bf case B}. }
\label{fig:7}
\end{center}
\end{figure}

This  case is a small variation of the previous one: the difference is
that now the left-handed quarks are localized towards the UV of the
quiver. The right-handed down sector would remain also in the UV, with
the right-handed up sector in the UV with one exception due to the
need for a large top quark mass.  This new solution is shown in
Figures~\ref{fig:5}, \ref{fig:6} and \ref{fig:7}. 
We see that the right-handed down sector remains universally coupled
to the gauge excitations, whereas now not only there is  flavor
violation in the right-handed up sector but also in the left-handed
sector, as shown in Figure~\ref{fig:5}. 
This results in slightly larger flavor violation in the down sector,
but still not as large as the effect in the continuum limit. 
In what follows we evaluate the amount of flavor violation incurred in
each case.  We will see that in both cases flavor violation bounds can
be satisfied. On the other hand, the difference between cases {\bf A} and {\bf B}
will be more relevant when computing electroweak precision
constraints. This will be done in the next section.

\subsection{Flavor Violation Bounds} 
In order to study the effects of flavor violation we need to obtain
the  tree-level couplings of fermions to the first-excited gauge
boson. Here we concentrate in quark couplings since they are the most constraining.
In both examples, cases {\bf A} and {\bf B},
Figures~(\ref{fig:2}--\ref{fig:7}) give the couplings of quarks to the first
gauge excitation in the interaction (diagonal) basis. To obtain the
flavor-violating interaction, we must rotate to the mass basis. 
For instance, if we define the couplings of the left-handed up-type quarks in the 
mass eigen-basis as given by
\begin{displaymath}
G_L^U \equiv  U_L^{-1}  \left(\begin{array}{ccc} g_{u_L} & 0 & 0 \\ 0 & g_{c_L} & 0  \\ 0 & 0 & g_{t_L}  \end{array} \right) U_L~,
\end{displaymath}
where $U_L$ is a unitary matrix that rotates to the mass eigenstates,
the couplings $g_{u_L}$, $g_{c_L}$ and  $g_{t_L}$ are the one given by
the dots in Figures~\ref{fig:2} and \ref{fig:5} for cases {\bf A} and
{\bf B}, respectively. 
The non-diagonal coupling matrix $G_L^U$ contains the flavor violating
couplings. 
There will be an analogous definition for the right-handed up sector
in terms of $U_R$, as well as for the down sector, both left- and right-handed. 
The tree-level flavor violation induced by the non-diagonal entries in
the $G$'s will result in bounds on the mass of the first excitation of
the gauge bosons in FHQT. In order to be conservative, we will
consider the effects of the first excitation of the gluon, that is we
assume that $SU(3)_c$ propagates in the quiver diagram. 
Although this is not necessary in these theories, it will provide the most
stringent bound and will allow us to compare with the analogous  flavor-violation
bounds in AdS$_5$, the continuum limit. 

The effective Hamiltonian for $\Delta F=2$ transitions will receive the
following contributions when the excited gluon is integrated out
\bear
H_{\rm eff.} &=& \frac{1}{M_G^2}\left[ \frac{1}{6} G^{ij}_L G^{ij}_L
  (\bar q_L^{i\alpha}\gamma^\mu q_{L\alpha}^j)  (\bar
  q_L^{i\beta}\gamma^\mu q_{L\beta}^j) + ({\rm L \leftrightarrow R})
\right.\nonumber\\
& & \left.-G_L^{ij}G_R^{ij}\left( (\bar q_R^{i\alpha}q_{L\alpha}^j)(\bar q_L^{i\beta}q_{R\beta}^j) 
- \frac{1}{3} (\bar q_R^{i\alpha}q_{L\beta}^j)(\bar q_L^{i\alpha}q_{R\beta}^j) 
\right)
\right]
\label{df2h}
\eear
This can be matched with the low energy $\Delta F=2$ Hamiltonian,
resulting in contributions to some of its Wilson
coefficients. Following Ref.~\cite{utfit}, the contributions from
(\ref{df2h}) result in 
\be
C^1_M(M_G) = \frac{1}{6}\frac{(G_L^{ij})^2}{M_G^2}, \qquad
C^4_M(M_G) = \frac{G_L^{ij}G_R^{ij}}{M_G^2}\qquad
C^5_M(M_G) = \frac{G_L^{ij}G_R^{ij}}{3 M_G^2}
\label{wilsoncs}
\ee
where $M=K,D,B_d,B_s$ refers to the particular meson. The Wilson
coefficients are then bound by the fits to flavor data by the UTFit
collaboration~\cite{utfit}, which provides the most comprehensive 
treatment of flavor data to bound new physics.  The bounds on these
coefficients from their latest fit of $\Delta F=2$ observables is
shown in the second column of Table~3. The bounds from kaon
physics were updated in Ref.~\cite{cecilia}. The third column gives
the bounds on the scale of new physics implied by assuming that the 
Wilson coefficients at $\Lambda$ are unity. In our case, the Wilson
coefficients given in (\ref{wilsoncs}) have a large suppression and
therefore  allow for a much lower energy scale $M_G$. Finally, given
that the bounds are obtained at larger scales $\Lambda$, in some cases
where this scale is considerably larger than $M_G$, we must correct
for the renormalization group evolution that is implied in the bounds.

We first study the bounds on {\bf case A}, presented in the previous
subsection and defined in Figures~\ref{fig:2}~to~\ref{fig:4}. This case is clearly designed to
minimize tree-level flavor violation effects. The choice of
localization for quarks in the quiver, defined by the values of the
various $c^i_{L,R}$ coefficients and consistent wit the physical masses
and with the observed $V_{\rm CKM}$, also fixes the non-diagonal
couplings of the mass-basis quarks to the first gauge excitation of
the gluon. In Table~1 we show the values of the non-diagonal
entries for $G_L^U$, $G_L^D$, $G_R^U$ and $G_R^D$. 
\begin{table}[h]
\begin{center}
        \begin{tabular}{|c|c|c|c|c|c|} \hline
         & L & R &  & L & R  \\ [1pt] \hline
         $|G^{u,c}|$ & $1.1\times10^{-5}$ & $2.2\times10^{-8}$ & $|G^{d,s}|$ & $5.7\times10^{-5}$ & $1.6\times10^{-9}$  \\ [1pt] \hline
         $|G^{u,t}|$ & $2.0\times10^{-4}$ & $2.3\times10^{-6}$ & $|G^{d,b}|$ & $1.9\times10^{-4}$ & $2.1\times10^{-8}$  \\ [1pt] \hline
          $|G^{c,t}|$ & $5.5\times10^{-6}$ & $6.8\times10^{-4}$ & $|G^{s,b}|$ & $5.9\times10^{-5}$ & $2.5\times10^{-6}$  \\ [1pt] \hline
                        \end{tabular}
\caption{{\bf Case A}: Non-diagonal values of the quark couplings to the first
  excitation of the gluon. }
\end{center}
\label{tab:1}
      \end{table} 
When comparing (\ref{wilsoncs}) with the third column of
Table~3, we see that for {\bf case A} there are virtually no meaningful
bounds coming from flavor physics on the mass scale of the first
excitations of gauge bosons, even if these are color-octect states. 
However, as we will see in the next section, the fermion localization
in the quiver chosen for {\bf case A}, will result in larger
contributions to the S and T  parameters. 

On the other hand, zero-mode fermion localization in {\bf case B} is
chosen so as to minimize effects on electroweak precision observables
(see next section). This results in larger flavor-violation
effects. The non-diagonal entries of the matrices  $G_L^U$, $G_L^D$,
$G_R^U$ and $G_R^D$ in {\bf case B}, which lead to tree-level flavor violation, are
given in Table~2. 
\begin{table}[h]\label{t:tabelax}
\begin{center}
        \begin{tabular}{|c|c|c|c|c|c|} \hline
         & L & R &  & L & R  \\ [1pt] \hline
         $|G^{u,c}|$ & $2.8\times10^{-3}$ & $2.9\times10^{-4}$ & $|G^{d,s}|$ & $5.7\times10^{-4}$ & $6.5\times10^{-6}$  \\ [1pt] \hline
         $|G^{u,t}|$ & $4.2\times10^{-3}$ & $2.9\times10^{-3}$ & $|G^{d,b}|$ & $5.9\times10^{-3}$ & $5.0\times10^{-5}$  \\ [1pt] \hline
          $|G^{c,t}|$ & $3.3\times10^{-2}$ & $1.8\times10^{-1}$ & $|G^{s,b}|$ & $6.7\times10^{-3}$ & $1.2\times10^{-4}$  \\ [1pt] \hline
                        \end{tabular}
\caption{{\bf Case B}: Non-diagonal values of the quark couplings to the first
  excitation of the gluon. }
\end{center}
      \end{table}
Comparing these entries with those of Table~1, we see that larger
flavor-violating effects are to be expected. These translate into
bounds on the mass of the gauge excitation. The bounds from $\Delta
F=2$ operators are shown in the last column of Table~3.  The most
constraining bound comes from the chirally-enhanced operator
$O_4$~\cite{cecilia}. The bound on $ImC_K^4$ results in 
\begin{equation}
M_G>3~{\rm TeV}~.
\label{flavbound}
\end{equation}
 Another bound from $\Delta S = 2$ physics that is close to this comes
 from $ImC_K^1$, which results in $M_G>2.6~$TeV. But a limit quite
 similar to that of (\ref{flavbound}) comes from charm physics. As can
 be seen in Table~3, the bound on $|C_D^4|$ requires $M_G>
 2.9~$TeV. This is due to the fact that the right-handed up sector must be more
 localized towards the $N$ site in order to get the correct top quark
 mass. In any case, the bound in (\ref{flavbound}) is just an example
 of a typical value that would pass all flavor limits in the
 particular solution
 shown in Figures~\ref{fig:5}, \ref{fig:6} and \ref{fig:7}. It is not
 necessarily the smallest possible value for $M_G$ for all
 solutions. However, it is a good illustration of the fact that in
 FHQT it is possible to get the fermion mass hierarchy without large
 flavor violation. This must be be contrasted with the typical flavor
 violation obtained in AdS$_5$ models, which results in much tighter bounds~\cite{csabafv}.
\begin{table}[!h]
\begin{center}
\begin{tabular}{@{}cccc}
\hline\hline
Parameter &~~ $95\%$ allowed range~~ &~~ Lower limit on $\Lambda$
(TeV)&~~ Bound on Color-octect\\
          & (GeV$^{-2}$) &for arbitrary NP & Mass in FHQT (TeV)\\
\hline
Re$C_K^1$ & $[-9.6,9.6] \cdot 10^{-13}$ & $1.0 \cdot 10^{3}$& $0.2$ \\
Re$C_K^4$ & $[-3.6,3.6] \cdot 10^{-15}$ & $17 \cdot 10^{3}$ & $0.1$\\
Re$C_K^5$ & $[-1.0,1.0] \cdot 10^{-14}$ & $10 \cdot 10^{3}$ & $0.1$\\
\hline
Im$C_K^1$ & $[-2.6,2.8] \cdot 10^{-15}$ & $1.9 \cdot 10^{4}$& $2.6$ \\
Im$C_K^4$ & $[-4.1,3.6] \cdot 10^{-18}$ & $49\cdot 10^{4}$ & $3.0$ \\
Im$C_K^5$ & $[-1.2,1.1] \cdot 10^{-17}$ & $29 \cdot 10^{4}$ & $1.0$ \\
\hline
\hline
$|C_{D}^1|$ & $<7.2 \cdot 10^{-13}$ & $1.2 \cdot 10^{3}$& $1.0$\\
$|C_{D}^4|$ & $<4.8 \cdot 10^{-14}$ & $4.6 \cdot 10^{3}$& $2.9$ \\
$|C_{D}^5|$ & $<4.8 \cdot 10^{-13}$ & $1.4 \cdot 10^{3}$& $0.5$\\
\hline
\hline
$|C_{B_d}^1|$ & $<2.3 \cdot 10^{-11}$ & $0.21 \cdot 10^{3}$& $0.3$\\
$|C_{B_d}^4|$ & $<2.1 \cdot 10^{-13}$ & $2.2 \cdot 10^{3}$ & $0.3$ \\
$|C_{B_d}^5|$ & $<6.0 \cdot 10^{-13}$ & $1.3 \cdot 10^{3}$ & $0.1$\\
\hline
\hline
$|C_{B_s}^1|$ & $<1.1 \cdot 10^{-9}$ & $30$ & $0.1$\\
$|C_{B_s}^4|$ & $<1.6 \cdot 10^{-11}$ & $250$& $0.1$\\
$|C_{B_s}^5|$ & $<4.5 \cdot 10^{-11}$ & $150$& $0.03$\\
\hline
\hline
\end{tabular}
\end{center}
\caption {$95\%$ probability range for
  $C(\Lambda)$ and the corresponding lower bounds on the NP
  scale $\Lambda$ for arbitrary NP flavor structure, from Refs~\cite{utfit,cecilia}.
The last column corresponds to the bound on the gluon excitation in
FHQT in {\bf case B} as described in the text.}
\label{tab:2}
\end{table}

In general, there will be a similar  situation for other values of $N$
as long as these are small enough to be away from the continuum
limit. Also, here we are studying FHQT with a very large cutoff, of
the order of the Planck scale. However, it is worth studying 
flavor models and their bounds with smaller UV cutoffs. We leave these
studies for future work.

\section{Electroweak Precision Constraints} 
\label{ewpc}
Another important set of bounds comes from precision electroweak
measurements. Particularly binding are the $S$ and $T$ parameters
defined as
\begin{equation}
S = 16\pi\left( \Pi'_{33}(0) -\Pi'_{3Q}(0)\right)
\label{sdef}
\end{equation}
and 
\begin{equation}
\alpha T = \frac{4}{v_{EW}^2} \left( \Pi_{11}(0) - \Pi_{33}(0)\right)
\label{tdef} 
\end{equation}
In order to compute $S$ and $T$, we first need to choose an
electroweak sector to propagate in the quiver.
The minimal choice is to take $G_j=SU(2)_L\times U(1)_Y$ for all
values of $j$. In AdS$_5$ this would be equivalent to have the SM
gauge fields propagate in the bulk. This is an unacceptable choice in
AdS$_5$ models since it results in too large isospin violation and
contributions to the $T$ parameter. As we will se below, this is not
the case for the coarse FHQT we study here, $N=4$. So we will study
the  electroweak sector of the SM propagating in the quiver.

\begin{figure} 
\begin{center}
\includegraphics[scale=0.9]{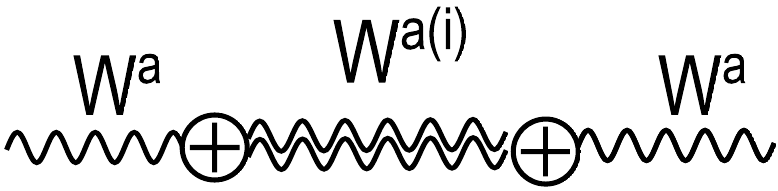}
\caption{
Diagrams contributing to $S$ and $T$ through the exchange of the
excited states of the $SU(2)L\times U(1)_L$ gauge bosons. The circled
crosses denote the mixing with the zero modes.}
\label{fig:8}
\end{center}
\end{figure}

\begin{figure} 
\begin{center}
\includegraphics[scale=0.9]{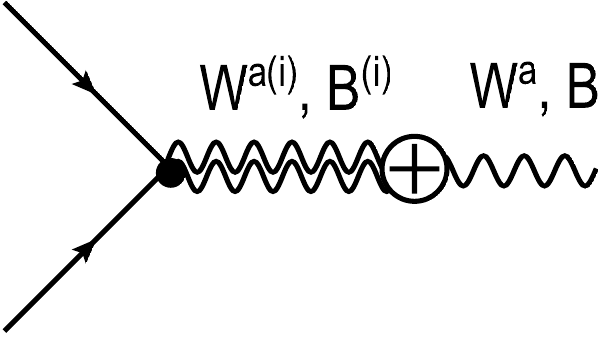}
\caption{
Universal tree-level renormalization of the $SU(2)_L\times U(1)_Y$
gauge couplings to fermions. The circled crosses denote the mixing of
the excited state with  the zero-mode gauge bosons. }
\label{fig:9}
\end{center}
\end{figure}

Although in most extensions of the SM contributions to $S$ and $T$
start arising at loop level, in FHQT -just as in AdS$_5$ models- there
are effects arising already at tree level. 
These are driven by the mixing of the low-lying gauge bosons $W ^\pm$
and $Z$ with their excited stated through the Higgs VEV.
The mixing effect leads to contributions to oblique parameters
through the exchange of excited states, as shown in
Figure~\ref{fig:8}.
However, these are not the
largest contributions to $S$ and $T$. There are additional contributions arising  
from the universal shifts in the gauge couplings of light fermions
which result from the mixing, as illustrated in Figure~\ref{fig:9}. 
The contributions to the $S$ parameter arising from diagrams like the ones in
Figure~\ref{fig:8} are suppressed by a factor of $v^4/M_1^4$ and will
not be the leading source of $S$, which receives much larger
contributions from the universal shift of gauge couplings that result
from the diagrams of Figure~\ref{fig:9}.
On the other hand, the
contributions of the exchange diagrams of Figure~\ref{fig:8} to $T$
cannot be neglected. 

As mentioned in the previous section, the choice of fermion
localization we called {\bf case A}, which has negligible flavor violation, will result in larger
contributions to $S$ and $T$.  This increase, with respect to 
{\bf case B}, comes from the fact that the universal shifts in gauge
couplings induced by the diagrams of Figure~\ref{fig:9} are enhanced
when the light left-handed fermions are localized close to the $N$
site.  We will then concentrate in the result for {\bf case B}, where
light fermions are localized closer to the UV site.  For this
localization 
the exchange diagrams of Figure~\ref{fig:9} result in   
\begin{equation}
T_e \simeq 0.05\times \left(\frac{3~{\rm TeV}}{M_1}\right)^2~,
\end{equation} 
where $M_1$ is the mass of the first excited gauge boson state.
The universal part of the vertex corrections illustrated in Figure~\ref{fig:9} can be
absorbed to a redefinition of the electroweak gauge fields, which
result in additional contributions to $S$ and $T$, analogously to the
AdS$_5$ case~\cite{adms}. The universal vertex corrections resulting
from the mixing of electroweak gauge bosons with their excited states
can be absorbed by the field redefinitions
\begin{eqnarray}
W^\pm &\to & W^\pm (1-g^2\delta) \nonumber\\
W^3 &\to& W^3(1-g^2\delta) +B\,g\,g'\,\delta\label{fieldshift}\\
B &\to& B(1-g'^2\delta) +W^3\,g\,g'\,\delta\nonumber~,
\end{eqnarray}
with 
\begin{equation}
\delta =   -\frac{g_{01}\, g}{4}
\,\left(\frac{v}{M_1}\right)^2\,f_{N,0}\,f_{N,1} ~.
\label{delta}
\end{equation}
In (\ref{delta}) $g_{01}$ is the coupling of the first excited state of
the gauge bosons to the zero-mode fermions, given  by~(\ref{g01def}). 
These shifts restore  the gauge couplings to their SM values at tree
level, but result in new contributions to oblique parameters:
\begin{eqnarray} 
S_v &=& 32\pi \delta \nonumber\\ 
T_v&=& \frac{8\pi}{\cos^2\theta_W}\,\delta~.
\label{stv}
\end{eqnarray}
(For comparison, the {\bf case A} localization results in vertex
contributions to $S$ and $T$  that are roughly a factor of four larger
than the ones in (\ref{stv}).)
Adding the exchange and vertex contributions we obtain, for the {\bf
  case B} localization
\begin{eqnarray}
S &\simeq& 0.17\times  \left(\frac{3~{\rm
      TeV}}{M_1}\right)^2\nonumber\\
T &\simeq &  0.16\times  \left(\frac{3~{\rm
      TeV}}{M_1}\right)^2\label{totalst}~.
\end{eqnarray}
The experimental fit to oblique parameters with $m_t=173~$GeV and
$m_h=126~$GeV as reference values, results in~\cite{gfitter}  $S^{\rm
  exp.}=0.03\pm0.10$ and $T^{\rm exp.} = 0.05\pm0.12$. 
Thus, we see that a mass scale of $M_1=3~{\rm TeV}$ is well within the 
$95\%$~C.L. bounds. Then, this is the bound on the excited gauge boson
masses that passes all flavor and electroweak tests. 

Additional contributions to $S$ and $T$ come from loops, particularly
of excited fermion states. We will study them elsewhere. There will
also be  contributions from the link field
sector~\cite{dischm}. These are sub-leading compared to the ones shown above.

\vskip0.5in
\section{Conclusions and Outlook }
\label{conc}
We have presented a class of theories in four dimensions, FHQT, which
have interesting properties for model building at the TeV
scale. Although these
theories can be obtained from the deconstruction of
AdS$_5$ models, they have distinct properties and phenomenology. 
They can accommodate large hierarchies, both between the IR and UV
scales, as well as in the fermion spectrum, just as AdS$_5$
models. However, as shown in Section~\ref{sec3}, they have less flavor
violation at tree level when compared with typical AdS$_5$ models,
even when fermion masses are entirely obtained from localization 
in the quiver and governed by order one parameters.
As a result, flavor bounds allow the IR mass scale to be lower than in
5D models, typically as low as $3~$TeV, without the introduction of
ad-hoc flavor symmetries. 
We also showed in Section~\ref{ewpc} that the  same localization models that have this
flavor bound, called {\bf case B} in
Section~\ref{sec3}, pass electroweak precision constraints for the
same value of the excited state mass scale, even without the
extension of the electroweak gauge sector to provide custodial protection.  This is quite different
from the AdS$_5$, where the bulk electroweak gauge sector must be
extended beyond the SM in order to avoid large contributions to the
$T$ parameter~\cite{adms}. Thus, despite being related to AdS$_5$
models by having them as their continuum limit, FHQT for small values
of N have distinct phenomenological features. In particular, it is
possible to build models of EWSB and fermion masses with them that
pass electroweak precision and flavor bounds while still having their
IR scale naturally close to the TeV scale.

There are several avenues to explore further. First, the Higgs sector 
must be introduced dynamically so as to naturally result in a light
Higgs when compared with the IR scale of a few~TeV. A way to achieve
this ``little hierarchy'' is for the Higgs to be a pNGB. This is very similar to the
composite Higgs models that are built in AdS$_5$~\cite{chm,chmgen}, and can
be essentially regarded as their (coarse)
deconstruction~\cite{dischm}.  Specific realizations of this idea in
the context of  FHQT will be presented elsewhere~\cite{higgs}.

The phenomenology of models of EWSB built using FHQT  should be
explored at the LHC, since  it is quantitatively   different from that
of AdS$_5$ models. For the case when color propagates in the quiver,
which was used as a way to obtain the strongest bounds from flavor
physics, production and decay of the first color-octect resonance at
the LHC is qualitatively similar to those of the KK gluon in
AdS$_5$. However, its couplings to zero-mode fermions are quite
different, as illustrated in Figures~\ref{fig:2} to \ref{fig:7}. For a
fixed value of the UV cutoff, these couplings depend on the number of
sites N. It will be necessary to do a detailed study of this
dependence in order to search for the resonances in specific FHQT. 
On the other hand, in general it is not necessary for color to
propagate in the quiver. The minimal models using FHQT will only contain the
electroweak sector. Thus the phenomenology of these weakly-coupled massive gauge 
bosons should be considered separately. Finally, the lepton sector of
FQHT must be studied to complete models of fermion masses, as well as
to study the complementary phenomenology with leptons in the final
state at the LHC. 

As the LHC increases its reach in the search for physics beyond the
SM, building models of the TeV scale with FHQT for  
EWSB and the Higgs sector, fermion masses and other questions typical
of this energy scale, will provide a rich phenomenology beyond
that of AdS$_5$ models. In particular, FHQT for small number of sites
N will give a more complete picture of these class of theories which
includes AdS$_5$ as a limit. Ultimately, and just as is the case  for AdS$_5$
models, the hope is that FHQT are a description of the 
 TeV  scale that will guide us through its
phenomenology to a deeper understanding of the underlying dynamics of
that scale.

\bigskip

{\bf Acknowledgments:}
The authors acknowledge the support of the State of S\~{a}o Paulo
Research Foundation (FAPESP), and the Brazilian  National Counsel
for Technological and Scientific Development (CNPq).



\end{document}